TABLE I

| Sim | Npart | $\Omega_b$ | b | c | $\rho_{crit}$ |
|-----|-------|------------|-----|-----|---------------|
| 1 | $64^3$ | 0.1 | 2.5 | - | - |
| 2 | $64^3$ | 0.1 | 2.5 | - | - |
| 3 | $64^3$ | 0.1 | 2.5 | 1 | $710^{-26}$ |
| 4 | $64^3$ | 0.1 | 2.5 | 3 | $710^{-26}$ |
| 5 | $64^3$ | 0.1 | 2.5 | 20 | $710^{-26}$ |
| 6 | $64^3$ | 0.1 | 2.5 | 100 | $710^{-26}$ |

TABLE II

| Sim | H | $M_b$ $10^{10}M_o$ | $M_T$ $10^{10}M_o$ | $M_b/M_T$ | $R_{Vir}$ (kpc) | $\lambda$ | $V_c$ (km/sec) | $K_{rot}/K_{tot}$ |
|---|---|---|---|---|---|---|---|---|
| 1 | 1 | 0.27 | 2.21 | 0.12 | 322 | 0.02 | 169 | 0.95 |
|   | 2 | 0.23 | 1.9  | 0.12 | 300 | 0.06 | 161 | 0.98 |

| Sim | H | $M_*$ $10^{10}M_o$ | $M_T$ $10^{10}M_o$ | $M_b/M_T$ | $R_{Vir}$ (kpc) | $\lambda$ | $V_c$ (km/sec) | $M_{gas}/M_{tot}$ |
|---|---|---|---|---|---|---|---|---|
| 3 | 1 | 0.22 | 2.19 | 0.12 | 317 | 0.02 | 168 | 0.027 |
|   | 2 | 0.17 | 1.93 | 0.12 | 317 | 0.05 | 161 | 0.03 |
| 4 | 1 | 0.22 | 2.19 | 0.12 | 317 | 0.02 | 168 | 0.03 |
|   | 2 | 0.15 | 1.79 | 0.12 | 309 | 0.04 | 158 | 0.04 |
| 6 | 1 | 0.27 | 2.21 | 0.12 | 322 | 0.02 | 169 | 0.06 |
|   | 2 | 0.2  | 1.9  | 0.11 | 305 | 0.06 | 162 | 0.085 |

# Analysis of Galaxy Formation with Hydrodynamics

Patricia B. Tissera[1,2]   Diego G. Lambas[3]   Mario Abadi[3]

March 21, 1996

Postal Address:
Astrophysics Department
The Blackett Laboratory, Imperial College
Prince Consort Road, London SW7 2BZ
United Kingdom

**Abstract**

We present a hydrodynamical code based on the Smooth Particle Hydrodynamics technique implemented in an AP3M code aimed at solving the hydrodynamical and gravitational equations in a cosmological frame. We analyze the ability of the code to reproduce standard tests and perform numerical simulations to study the formation of galaxies in a typical region of a CDM model. These numerical simulations include gas and dark matter particles and take into account physical processes such as shock waves, radiative cooling, and a simplified model of star formation. Several observed properties of normal galaxies such as $M_{gas}/M_{total}$ ratios, the luminosity function and the Tully-Fisher relation are analyzed within the limits imposed by numerical resolution.

Subject Headings: Galaxy Formation; Cosmology; Hydrodynamical Simulations

1 Nuclear and Physics Laboratory, University of Oxford, United Kingdom

2 Actual addresses: Universidad Autónoma de Madrid, Departamento de Física Teórica C-XI, Cantoblanco, Madrid 28040, Spain; Imperial College of Science, Technology and Medicine, The Blackett Laboratory, Prince Consort Road, London SW7 2BZ, United Kingdom

3 Observatorio Astronómico de la Universidad Nacional de Córdoba, Laprida 853, 5000 Córdoba, Argentina

# 1 Introduction

The understanding of the physical processes involved in the formation and evolution of the structure in the Universe has significantly improved over the last years. The observational data available at present provide important information about the properties of the structure at different redshifts, encouraging the performance of more elaborated semi-analytical and numerical models. In particular, N-body simulations of hierarchical clustering models have allowed the studies of a great variety of dynamical phenomena associated with galaxies and larger structures. In these simulations, a more consistent handling of both, highly non-linear processes and large dynamic range have been possible through the development of adequate numerical techniques and fast computers.

Numerical simulations of collisionless particles give a reliable description of the gravitational evolution of the dark matter component and have been significantly useful for studying the formation of structure in different cosmological models (Frenk et al. 1988; Davis et al. 1985; White & Frenk 1991; Tissera et al. 1994). Although in these simulations, the behavior of the dissipative component is derived through simple modelling based on phenomelogy, they have provided important clues for understanding the physical processes involved in the evolution of the structure. In order to overcome some of the shortcomings of these models, hydrodynamics has been incorporated to purely gravitational codes by several authors ( Evrard 1988, Hernquist & Katz 1989, Cen et al. 1990; Summers 1993; Navarro & White 1993). Astrophysical problems require the use of numerical techniques to account for hydrodynamic effects due to the fact that analytical approaches are restricted to systems with special symmetries. These techniques are a powerful tool for the study of the fully non-linear hierarchical clustering when a gaseous component is present.

Smooth Particle Hydrodynamics (SPH) is a widely used numerical method for the treatment of gas dynamics. It was first introduced by Gingold and Monaghan (1977), and Lucy (1977) for the study of stellar pulsation, and in the last years, has SPH technique been applied to cosmological studies. SPH has been previously implemented and tested in Tree, PM and P3M algorithm (Evrard 1988; Hernquist & Katz 1989; Navarro & Benz 1991; Thomas & Couchman 1994). These implementations have provided the first insights in a more consistent treatment of systems with gas and dark matter.

Dissipative effects play a critical role in the evolution of the structure in galactic scales. They strongly affect the dynamical behavoir of the bary-onic component making possible the formation of rotational supported disks (Katz 1992; Navarro & White 1994). Therefore, a complete picture of galaxy formation requires the understanding of the accretion process of baryons into the potential well of the halos. On the other hand, the hydrodynamical treatment of the dissipative component provides a more consistent framework to implement models for star formation. Star formation is a critical process that influences many aspects of the formation and evolution of galaxies, although a fully consistent treatment of these processes has not been yet achieved. A reliable model for star formation in numerical simulations is difficult to achieve due to our poor understanding of the physics involved and the limits imposed by numerical resolution. There have been several attempts of modelling star formation processes in SPH numerical simulations (Katz 1992; Navarro & White 1993). These authors have introduced schematic algorithms for transforming the cold gas present in highly dense clumps into stars. Albeit simple, these algorithms provide a basic framework for the assessment of the effects of star formation processes on the properties of the galaxies, and from which we can underpin a more realistic model.

In this paper, we develop an implementation of SPH technique which follows the outline described by Hernquist and Katz (1989) into the Adaptative Particle[3] Mesh code (AP3M) developed by H. Couchman (1992). In Section 2, we present a brief description of SPH and the implementation performed in this paper. Section 3 describes the tests of the code. In Section 4, we study the formation of galaxies in a typical region of the Universe without considering star formation. In Section 5, we introduce a simple star formation scheme and analyze the astrophysical properties of the resulting galactic-sized objects. Section 6 outlines the main conclusions achieved.

## 2  SPH Implementation

SPH is a purely Lagrangian method which allows the numerical integration of the hydrodynamical equations of a continuous fluid. The fluid distribution is approximated by a set of N-particles each one representing a differential fluid volume element. Since the number of volume elements is finite, local averages are required to estimate the value of a physical variable at a particle point. In SPH, the mean value of a physical field $f(r)$ is estimated according to:

$$\langle f(r) \rangle = \int W(r - r_j, h) f(r_j) dr_j \qquad (1)$$

where $W$ is the smoothing kernel and $h$, the smoothing length. We assume the approximation $f(r) = <f(r)>$ which has associated errors of order $h^2$. It can be easly proved that the smoothed value of a spatial derivate of a function can be constructed based on the spatial derivates of the kernel. This fact makes simple and straighforward the assessment of any physical variable and its derivates (see Benz 1990; Hernquist & Katz 1989 for more details).

In a dicrete medium where the values of $f(r)$ are known only at a finite number of points with number density $n(r) = \sum_{b=1}^{N} \delta(r - r_j)$, the smoothed value of $f(r)$ can be estimated as follows:

$$\langle f(r) \rangle = \sum_{b=1}^{N} \frac{m_b}{\rho_b} f(r_b) W(r - r_b, h) \qquad (2)$$

where the sum is performed over all neighbour particles with mass $m_b$ and density $\rho_b$.

In order to work with symmetrized functions, we use an average of kernels (Hernquist & Katz 1989) defined by :

$$f(r_a) = \sum_{b=1}^{N} \frac{m_b}{\rho_b} f(r_b) \frac{1}{2} \left[ W(r_a - r_b, h_a) + W(r_a - r_b, h_b) \right] \qquad (3)$$

The notation $W_{ab} = \frac{1}{2} \left[ W(r_a - r_b, h_a) + W(r_a - r_b, h_b) \right]$ will be used henceforth.

The smoothing length $h(r, t)$ and the shape of the kernel determine the volume which effectively contributes to the smoothed value of $f(r)$. In this work, we adopt a kernel that corresponds to a spline function (Monaghan & Latanzio 1985):

$$W(r, h) = \frac{1}{\pi h^3} \left[ \begin{array}{ll} 1 - (3/2)(r/h)^2 + (3/4)(r/h)^3 & 0 < r/h < 1 \\ (1/4)[2 - (r/h)]^3 & 1 < r/h < 2 \\ 0 & r/h > 2 \end{array} \right] \qquad (4)$$

The time evolution of the smoothing length can be computed assuming that it scales with the interparticle distance, and resorting to the continuity equation to replace the temporal derivate by spatial ones which can be simply expressed in the SPH formalism (Evrard 1988):

$$\frac{dh_a}{dt} = -\frac{h}{3\rho}\sum_{b=1}^{N}(\vec{v}_a - \vec{v}_b)\cdot\nabla W_{ab}(r_a - r_b, h_{ab}) \qquad (5)$$

We calculate the number of neighboring particles ($N_{vec}$) within a spherical volume of radius $2h_a$ for all particles using the estimated value of the smoothing length given by Equation 5. In order to keep the local character of the variables throughout the simulation, and calculate them with the same precision all over the integration volume, we require that all particles have a number of neighbours between 30-50 according to previous works (see for instance Navarro & White 1993). Therefore, if this condition is not satisfied we correct the smoothing lengths applying to the algorithm proposed by Hernquist and Katz (1989):

$$h_a^{n+1} = \frac{h_a^n}{2}[1 + (N_{vec}/N_{med})^{1/3}] \qquad (6)$$

where $N_{med}$ is a parameter fixed to 40. We impose a lower limit equal to half the gravitational softening to the smoothing lengths (Summers 1994; Navarro & White 1993).

We assume that the baryonic component is initially in the form of ideal gas with internal energy $\epsilon = kT/(\gamma - 1)\mu m_p$ (where $\mu$ is the mean molecular weight, $m_p$ the proton mass and $k$ Boltzmann's contant). The evolution of $\epsilon$ is specified by the energy conservation equation with the addition of an extra term $\Lambda(\epsilon)$ that takes into account the dissipation of energy due to radiative cooling:

$$\frac{d\epsilon}{dt} = \frac{P}{\rho^2}\frac{d\rho}{dt} - \frac{\rho\Lambda(\epsilon)}{(\mu m_p)^2} \qquad (7)$$

where $P = (\gamma - 1)\rho\epsilon$ is the gas pressure.

Applying the SPH formalism to equation (7) results in:

$$\frac{d\epsilon_a}{dt} = \sum_b m_b(\frac{P_a}{\rho_a^2} + \Pi_{ab})v_{ab}\cdot\nabla W_{ab}(r_a - r_b, h_{ab}) - \frac{\rho\Lambda(\epsilon_a)}{(\mu m_p)^2} \qquad (8)$$

where the function $\Pi_{ab}$ represent an artificial viscosity force and the notation $v_{ab} = (v_a - v_b)$ has been used.

This artificial viscosity force has been introduced in order to accomplish a numerically correct description of the evolution of the fluid in a discontinuity layer (Gingold & Monaghan 1977). In a real fluid with a high Mach number where shocks may develop, the viscosity produces the transformation of kinetic energy into heat. The inclusion of artificial viscosity forces makes possible to resolve numerically the shocks, generating the correct exchange of energies and damping non-physical oscillations on the scale of particle separation (see Shu 1992). We have adopted the following functional form proposed by Gingold and Monaghan:

$$\Pi_{ab} = \begin{cases} \frac{-\alpha c_{ab} \mu_{ab} + \beta \mu_{ab}^2}{\rho_{ab}} & \vec{v} \cdot \vec{r} \leq 0 \\ 0 & \text{otherwise} \end{cases}$$

where

$$\mu_{ab} = \frac{h \vec{v}_{ab} \cdot \vec{r}_{ab}}{r_{ab}^2 + \eta h^2} \tag{9}$$

In order to compute the effects of the radiative cooling, we adopt the approximation for $\Lambda(\epsilon)$ given by Dalgarno & McCray (1972), for a primordial mixture of hydrogen and helium (He/H=0.25) (Katz 1992). In the case where dissipative processes are present, the integration of (7) demands extra caution. Taking into account this fact, we have updated the internal energy $\epsilon$ using a semi-implicited integration method based on a predictor-corrector scheme (Hernquist & Katz 1989). Since the AP3M code uses the same time step for all particles, this semi-implicited method allows to follow more carefully the integration of the internal energy. This is particularly important for particles situated in high density regions where cooling timescales are shorter than dynamical timescales.

The complete set of equations of motion in comoving coordinates ($\vec{x}$) is:

$$\frac{d\vec{x}}{dt} = \vec{v} \tag{10}$$

$$\frac{d\vec{v}}{dt} = -2H\vec{v} - \frac{1}{a^2}\frac{\nabla P}{\rho} - \frac{1}{a^2}\nabla \Phi(x) \tag{11}$$

where the gradient of the pressure in the SPH formalism is estimated as:

$$\frac{\nabla P}{\rho} = -\sum_b m_b (\frac{P_a}{\rho_a^2} + \frac{P_b}{\rho_b^2} + \Pi_{ab})\nabla_{ab} W(r_a - r_b, h_{ab}) \quad (12)$$

We have implemented the contribution of hydrodynamical forces into the AP3M code using a standard leapfrog algorithm to update particle positions and velocities.

The cosmic energy equation for a fluid in comoving coordinates (Peebles 1980) modified to take into account the gas pressure and radiative energy losses is used for checking the performance of the integration throughout the simulations.

## 3 Self-similarity solutions for an adiabatic spherical collapse

The reproduction of the self-similarity solutions of an adiabatic spherical collapse has become a classic test for hydrodynamical codes in three dimensions. Bertschinger (1985) studied the evolution of a spherical perturbation in an otherwise uniform Einstein-de Sitter Universe model, finding the following self-similarity solutions for the evolution of the pressure ($p$), density ($\rho$), mass ($m$) and velocities ($v$) fields in the secondary accretion collapse :

$$v(r,t) = \frac{r_{ta}}{t} V(\lambda) \quad (13)$$

$$\rho(r,t) = \rho_H D(\lambda) \quad (14)$$

$$p(r,t) = \rho_H (\frac{r_{ta}}{t})^2 P(\lambda) \quad (15)$$

$$m(r,t) = \frac{4}{3}\pi \rho_H r_{ta}^3 M(\lambda) \quad (16)$$

where

$$\rho_H = (6\pi G t^2)^{-1} \quad (17)$$

$$r_{ta}(t) = (\frac{3\pi}{4})^{-8/9} \delta_i^{1/3} r_i \tau^{8/9} \quad (18)$$

$$\lambda = \frac{r(t)}{r_{ta}(t)} \quad (19)$$

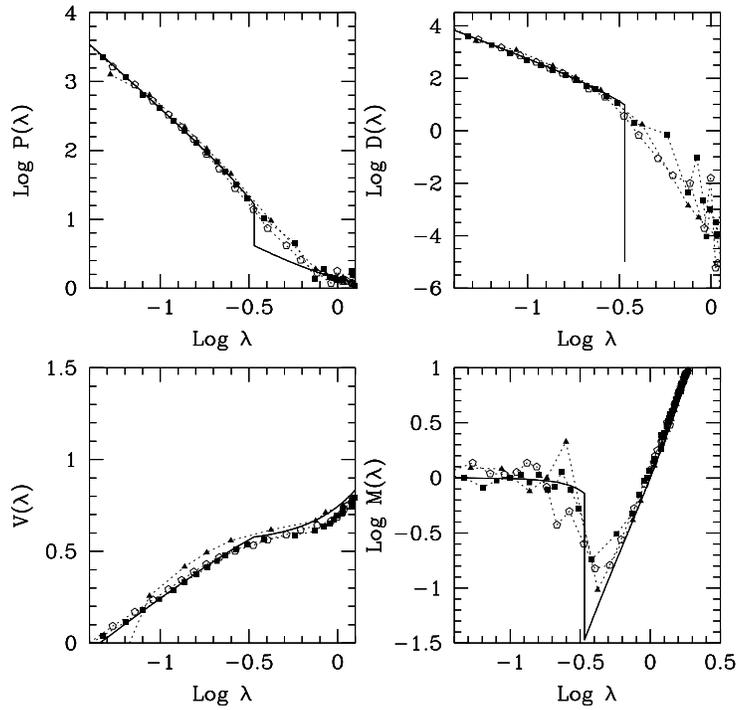

Figure 1: Self-Similarity solutions for an adiabatic spherical collapse for density ($D$), pressure ($P$), mass ($M$) and velocity ($V$) fields. Simulated solutions for different particle numbers inside the $r_{shock}$ are shown: $N > 150$ (filled triangles), $N > 200$ (open pentagon) and $N > 300$ (filled squares). Analytical solutions are shown for comparison (solid line).

We have simulated the collapse of a spherical adiabatic perturbation of a collisional fluid in order to compare the results of our code with these semi-analytic solutions. We have followed the accretion of an initial uniform distribution of N=4096 gas particles onto a central perturbation with a density contrast $\delta\rho/\rho \simeq 5$, and computed the parametrized functions $V(\lambda), D(\lambda), P(\lambda)$ and $M(\lambda)$ at different stages of the evolution.

Figure 1 shows the analytical solutions found by Berstchinger (solid lines) and the relations obtained from the simulations at different times that corre-spond to $N_{gas} = 139, 199, 420$ particles inside the shock radius ($r_s = 0.34 r_{ta}$). As we can seen in this Figure, the fit to the theoretical solutions improves as the number of particles inside the radius shock increases. These results are sastisfactory and assure that the code is able to correctly reproduce the behaviour of the gaseous component through and after a discontinuity layer.

# 4 CDM Numerical Simulations

In this section, we study the formation of galactic objects in a typical region of a Cold Dark Matter (CDM) model following the evolution of dark and baryonic matter. We analyze two different models. Model A considers the gravitational and hydrodynamic evolution of the matter including the treatment of pressure, shocks and radiative cooling, whereas Model B also introduces a schematic algorithm to take into account the effects of star formation processes.

The initial conditions of the simulations correspond to a biased CDM power spectrum with a density parameter $\Omega = 1$, a zero cosmological constant $\Lambda = 0$ (we adopt a Hubble's constant $H = 50\ km/s/Mpc$). We set initial positions and velocities for $N = 64^3$ particles according to the Zel'dovich approximation in a computational box of $L = 10\ Mpc$ and $64^3$ grids with periodic boundary conditions. Dark and gas particles have equal mass $m = 2.610^8\ M_\odot$. Gas particles are selected at random from the initial conditions according to a baryon density parameter $\Omega_b = 0.1$. The simulations were evolved until the rms mass fluctuations reached a value of 0.4 in a sphere of radius 16 $Mpc$ (bias parameter $b = 2.5$). The softening length for the gravitational forces is $\simeq 5$ kpc fixed in physical units. Table I summarizes the main characteristics of the simulations perfomed.

## 4.1 Model A without star formation

We first analyze simulations without including star formation processes. This model is a first approach to study the formation of structure on galactic scales due to both our rudimentary understanding of the physics involved and the complexity of its numerical implementation.

In Figures 2 are shown the projected distributions of the dark matter (Figure 2a) and the baryonic particles (Figure 2b) at redshift ($z = 0$) for a subregion of size $L \simeq 3\ Mpc$ of simulation 1. Due to the high efficiency of the radiative cooling, most of the gas in high density regions remains at the minimum allowed temperature $T \simeq 10^4 K$. In Figure 2b, it can be seen that the cold gas settles into clumps which are easily individualized. If we look more carefully at those small clumps, disk-like structures may be identified. Figures 2 c,d show the projected distribution of particles corresponding to a zoomed region of 500 $kpc$ and 150 $kpc$ box size respectively. Although our dynamical resolution range is not as high as desired, the formation of disk systems can be followed quite well in massive objects ($M_{gas} > 410^{10}$).

Figure 2: Projected distribution of dark matter (a) and gas (b) particles in a subsample of L=3 Mpc of simulation 1; c) and d) are closer views of the subregions delimited by the boxes.

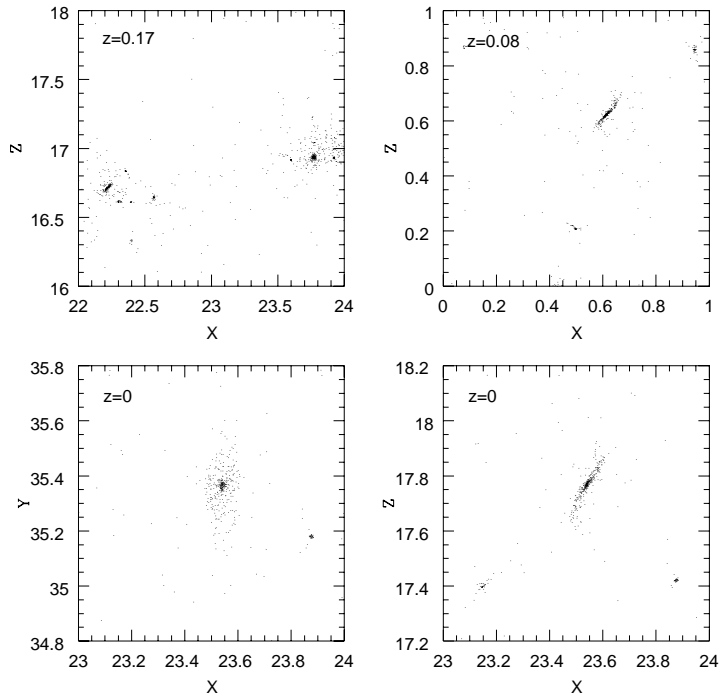

Figure 3: History of formation of Halo 3. a) Two colliding disks at z=0.18. b) The result of the merger at z=0.8. c) and d) Final configuration of disc at z=0 (planes xy and xz, respectively).

We have identified galactic halos applying an algorithm based on a density contrast criteria, $\delta\rho/\rho \propto 200$ (White et al 1991; Lacey et al 1993). Table II summarizes the principal characteristics of two clumps, halo 1 and 2, which will be analyzed in detail henceforth.

We traced back in time the evolution of the particles belonging to halos identified at z=0 in order to study their formation history. Disk-like structures form as the result of the hierarchical aggregation of substructure and mergers with other objects. As an example, Figures 3 a, b and c show the evolution of gas particles belonging to halo 2. We can see in these figures two disk objects at $z = 0.17$, the result of their merger at $z \simeq 0.08$ and the final configuration at z=0.

This example shows that gaseous disks can be regenerated after subse-quent mergers of initial disk-like objects as pointed out by Navarro & White (1994). We also observe that when a merger occurs the compact cold gas clumps orbit around each other before the merger effectively happens. It can be seen that several galaxy-like objects in our simulations have satellites neighbours inside $r_{200}$ as the result of recent mergers ( $r_{200}$ is the radius where the density contrast satisfied the condition $\delta\rho/\rho \propto 200$).

We have calculated the ratio between the rotational kinetic energy and the total kinetic energy of the disk structures obtaining typical values greater than 0.5. In particular, Halos 1 and 2 have $K_{rot}/K_{tot} \simeq 1$ as shown in Table II, indicating that these disk systems are rotationally supported. On the other hand, we have estimate the ratio between baryonic mass to total mass inside the virialized region of halos ($r < r_{200}$) $M_b/M_{vir} \simeq 0.11 - 0.12$ which is consistent with the mean baryonic density parameter $\Omega_b = 0.1$, albeit slightly greater. We find that the cold gas mass is more concentrated than dark matter reaching comparable values at a radius $r \simeq 10 - 15\ kpc$.

According to White & Frenk (1991) the virial mass $M_{vir}$ of a halo formed at redshift z can be expressed as a function of its circular velocity $V_c$:

$$M_{vir} = 2.35(1+z)^{-3/2}10^5 V_c^3 \tag{20}$$

In our simulations, we find that the gas settled in halos also satisfies this relation ($M_{gas} \propto V_c$). It should be remarked that the total cold gas mass in the simulated halos exceeds by a factor of 5 the total observed stellar mass content in galaxies. So, if it were assumed that the total accreted gas mass in the simulated halos were gradually transformed into stars, the resulting stellar content would be in open disagreement with observations. This problem may be overcome introducing heating sources like supernovas which may prevent the gas to overcool and settle into the potential well of the halos from early times.

In order to explore the possible presence of velocity bias in the hydrodynamical simulations, we have computed at different redshift the ratio of the rms velocity dispersion of halos $<v_{halo}^2>$ and mass $<v_m^2>$, $b_v^2 = \frac{<v_{halo}^2>}{<v_m^2>}$. The results of these calculations show no significant velocity bias ($b_v \simeq 1$).

## 4.2 Models B including star formation

Star formation is a critical factor in the evolution of the galactic structure. The physical processes involved are a key point in the study of galaxy formation although they are still not well understood. Moreover, it is difficult to model them realistically taking into account the current computational resolution ranges. In this section, we introduce a simple star formation (SF) algorithm but for reasons of simplicity we have not considered stellar feedback processes.

Our prescription for SF is based on the model described by Navarro & White (1993). Following these authors, a gas particle is assumed to be in condition of forming stars if the Jeans inestability condition is satisfied and if its cooling time is shorter than its dynamical time. A critical density ($\rho_{crit} \simeq 7.10^{-26} gr/cm^3$) can be estimated such that if $\rho_{gas} \geq \rho_{crit}$, the second condition is satisfied. It is assumed that the Jeans inestability condition if fullfilled if the gas particle is in a convergent flow ($\nabla \cdot \vec{v} < 0$).

Once a particle representing a gas cloud satisfies these requirements, it is transformed into stars according to a certain star formation rate (SFR). We assume the following SFR:

$$\frac{d\rho_{gas}}{dt} = -c\frac{\rho_{gas}}{t_*} \qquad (21)$$

where $c$ is a dimensionless star formation rate parameter and $t_*$ is the characteristic time of the star formation process (Katz 1992). We assume that a gas particle is transformed instantaneously into stars after a time interval ($\tau$) over which the gas mass is expected to be transformed into stars according to Equation 21 (Navarro & White 1994).

### 4.2.1 Analysis

In order to analyze the effects of the star formation efficiency, we have performed several simulations sharing the same intial conditions and with different values of the efficiency parameter $c$ (Table II). The SF scheme described above may be very efficient in transforming the gas into stars. This high efficiency could lead to spurious results if stars are formed at early stages of the evolution, producing an excess of stellar mass and inhibiting the formation of disk structures.

We identified galactic halos at different redshifts using the algorithm previously mentioned. Table II summarizes the main characteristics of the halos 1 and 2 at $z = 0$. The detailed analysis of the properties of galactic halos will be considered in a forthcoming paper (Tissera 1996, In preparetion) As a consequence of our hydrodynamical resolution range, the evolution of the gas inside halos under $V_c \simeq 50 km/sec$ (or $M_{vir} \leq 10^{11}$) can be not well represented. This fact may delay artifitially the starting of the process of star formation. The maximum star formation rate is reached on average at $z \simeq 1$. At this redshift approximatly 8% of the total baryonic mass has been transformed into stars. This fraction increases to $15 - 30\%$ at $z = 0$, producing a baryonic density parameter estimated from stars in simulated galaxies of $\Omega_* \sim 0.015 - 0.03$, while the observed value is $\Omega_* \simeq 0.004$ (Peebles 1993). This significant excess of stars can be related to the overcooling effect discussed by several authors (Cole 1994; Navarro & White 1994) which generates an early collapse of the gas into the clumps. On average, the total cold gas mass in the simulated galaxies at $z = 0$ is approximatly $45 - 50\%$ of the total baryonic mass in the volume considered. Heating processes like supernova explosions or photoionization might prevent the gas from over collapsing into small dense cold clumps reducing the amount of matter in condition of forming stars. This effect would also be enhanced in hierarchical models due to the over abundance of small halos (White & Frenk 1991; Cole 1994)

The different efficiency parameters $c$ considered generate different histories of star formation implying differences in the astrophysical properties of the final objects. A high SF efficiency can tranform the gas into stars rapidly enough to produce spheroidal structures. An adequate adjusting of this parameter allows the gas to settle into a disk-like structure while being transformed into stars. Figure 4 shows the distributions of dark, gas and star components of halo 3 of simulation 6 where a gaseous disk and a budge of stars could be formed with a particular value of $c$.

In simulation 6 the maximum of the SF rate is reached at a low redshift ($z \simeq 0.3$). This delay of the SF processes, implies a more recent SF history. On the other hand, in the models analyzed, massive disks settle at $z \geq 0.5$ on average, fact that agrees with the conclusions found by Navarro & White (1993). This result may be affected by several factors including numerical resolution, so more detailed simulations are required to address more consistently this problem.

### 4.2.2 Astrophysical Properties

We estimate the luminosities of simulated galaxies in different bands taking into account the stellar evolutionary tracks computed by Bruzual &

Figure 4: Projected distribution of dark matter (a), gas (b) and stellar mass (c) for Halo 3 at z=0 in simulation 5.

Charlot (1993). These analytical tracks give the evolution of the luminosity per unit mass as a function of age for a given IMF and mass-to-luminosity ratio. For each star particle, we compute the corresponding luminosities at each time step taking into account its formation redshift. We derive the total luminosities of the halos adding the individual luminosities of each star particle inside $r_{200}$.

The Tully-Fisher relation for galaxies also provides a confrontation between models and observations. Several authors have found fits to this relation in different bands (see for instance Pierce and Tully 1992). It is observed that slope and the zero point of the fits to the TF relation may vary over an important range of possible values.

As a first step to understand the physics involved in the Tully-Fisher relation, we have analyzed the dependence of the total baryonic mass and the stellar mass on circular velocity. We considered two different subsamples according to the total baryonic particle number inside $r_{200}$ in order to analyze the effects of the numerical resolution. According to Navarro & White (1993), the collapse of a halo can be reliably computed with a minimum of $N_{gas} \simeq 300$ particles. Consequently, we consider two subsamples of halos: subsample I with $N_{gas} \geq 300$, and subsample II with $N_{gas} \geq 30$. We computed the mass in stars ( $M_*$ ) of the halos as a function of circular velocity ($V_c$) for both subsamples. For subsample I, we obtain $M_* \propto V_c^{2.8\pm0.3}$, and for subsample II, $M_* \propto V_c^{3.3\pm0.2}$. According to White & Frenk (1991) the total virial mass and the circular velocity in galactic halos satisfy $M_{vir} \propto V_c^3$. We find that the stellar content is also proportional to $V_c^3$ suggesting that star formation does not affect strongly the accretion of baryonic mass into the potential well of the halos.

We estimated the Tully-Fisher relation for the simulated galaxies in the blue and infrared band at $z = 0$. As it can be seen in Figure 5, the slopes of the relations derived from the simulations are in a reasonable agreement with the plotted observed values taking into account the observational error bands. Both simulated relations have been rescaled to match the observed zero points. It has to be pointed out that objects with $V_c < 50$ km/s have not been included.

## 5 Conclusions

The aim of this paper is to present this hydrodynamical code. This code allows us to study the evolution of structure in a cosmological context in-cluding a correct treatment of gas pressure, shocks and radiative cooling, as well as introducing a simplified treatment for star formation. The code has been tested and results compared with others from existing codes. Although improvement has to be done in order to increase its dynamical resolution range, the evolution of the structure and the study of their global astrophysical properties can be perfomed significantly well.

Figure 5: Tully Fisher (TF) relation in the blue (triangles) and infrared (pentagons) bands for halos in simulation 3. The corresponding observed TF (Piece & Tully 1992) are shown for comparison (dash line).

We also analyzed the formation of structure in a typical region of a CDM universe looking for hints to understand the physical processes involved and to assess the performance of the code in this sort of problem. ¿From the analysis of the set of simulations perfomed, we may conclude the following:

1- In agreement with previous works, we find that most of the gas settles into cold dense clumps that in the absence of heating sources, remains at $T \propto 10^4 K$. The total cold gas mass in halos exceeds the total stellar content observed in core galaxies suggesting the need of introducing heating sources to reduce the overabundance of cold gas clumps, and also to produce the amount of hot gas observed.

2- Disks are formed through the merger of substructure and on average in our simulations, they settle at late redshifts ($z \simeq 0.5$). Mergers are strongly important in the evolution of galactic halos given their high frequency. It is observed in our simulations that, in general, disks can be regenarated after a merger.

3- In models B we introduce a simple star formation scheme as a first step towards building up a more realitic model of galaxy formation. The inclusion of SF allows us to estimate the luminosities of the objects. Through an adequate adjusting of SFR efficiency we reproduce the observed ratio $M_{gas}/M_{total}$. However, the total amount of stellar mass in galaxies found in the simulations exceeds the observational value, resulting in a stellar density paramenter 3-5 times greater than the observed one.

4- When SF processes are included, the shape of the objects depends strongly on the efficiency parameter $c$ chosen. For appropiate values we recover disk-like objects but at the price of shifting the star formation to more advance redshifts, in disagreement with observations.

5- The baryonic mass accreted into the potential well of a halo is proportional to its total virialized mass although slightly higher than $\Omega_b M_{vir}$. This result is not affected when star formation is included and the relation $M_* \propto V_c^3$ remains valid.

6- The TF relation is marginal fit in the infrared and blue band depending on the observational constrains adopted. Supernovas may help to achieve a better fit at lower circular velocities, althought the relation $M_*$ vs $V_c$ should not be strongly affected.

These results are the first outcomings from the application of this version of AP3MSPH to the study of galaxy formation in a cosmological context. There are several points which require to be addressed more exhaustively. Future improvements of the code will permit to increase the numerical resolution of the simulations enabling a better description of the joint evolution of the baryonic and dark matter.

## 6 Acknowledgements

The authors would like to thank G. Efstathiou, S.D.M. White, J. Navarro, C.Frenk and B.Jones for usefull discussions and suggestions. We are gratefull to the British Council, CONICOR, CONICET, the University of Cordoba, the Univestiy of Durham and the Univestiy of Cambridge for their support. P.B.Tissera thanks Dr.George Efstathiou for his support and unvaluable suggestions and comments wich made possible the accomplishement of her Ph.D thesis and this paper. P.B.Tissera thanks the Astrophysics Department of the Univesity of Oxford and all its members for providing a place and facilities to support her research.